\documentclass[pdflatex,sn-mathphys-num]{sn-jnl}

\usepackage{graphicx}
\usepackage{amsmath,amssymb,amsfonts}
\usepackage{amsthm}
\usepackage{mathrsfs}
\usepackage{xcolor}
\usepackage{textcomp}
\usepackage{booktabs}

\theoremstyle{thmstyleone}

\theoremstyle{thmstyletwo}

\theoremstyle{thmstylethree}

\begin{document}

\title{Variational Openness: An Open Formulation of Hamilton's Principle}

\author*[1]{\fnm{Francisco} \sur{Monroy}}

\affil*[1]{
\orgdiv{Departamento de Química Física},
\orgname{Universidad Complutense de Madrid},
\orgaddress{
\street{Av. Complutense s/n},
\postcode{28040},
\city{Madrid},
\country{Spain}
}
}

\abstract{
Since its classical origin, Hamilton's principle has been formulated under an exact closure condition: admissible variations vanish at the boundaries of the variational domain. This condition removes the boundary term in the first variation of the action and yields the Euler--Lagrange equation. Although natural for isolated deterministic systems, fixed boundary admissibility is usually treated as a technical condition rather than as a physical closure hypothesis. Here we ask what follows when this hypothesis is made explicit and relaxed. We introduce \emph{variational openness} as the retention of the boundary contribution in the variational balance. The retained term defines a boundary-openness density, which must be projected onto admissible variations before it becomes a dynamical source. In this formulation, the classical Euler--Lagrange equation is recovered as the exact-closure limit of an open variational balance; the source term is therefore identified with incomplete variational closure rather than with an externally imposed force. The framework is illustrated through three elementary examples: an open harmonic oscillator, a finite-compliance boundary, and a delayed oscillator with memory. These examples show how boundary openness can produce forcing, partial closure, history dependence, and non-Markovian structure while preserving standard mechanics in the closed limit. The resulting perspective suggests that Hamiltonian mechanics may be understood as the mechanics of variationally closed systems and motivates an open Hamilton--Jacobi theory in which admissibility itself becomes dynamical.
}


\maketitle

\section{Introduction}
\label{sec:introduction}

Hamilton's principle is a cornerstone of theoretical physics. Introduced by Hamilton in the nineteenth century \cite{Hamilton1834}, it provides one of the most compact formulations of classical dynamics. Jacobi's later formulation made the action itself the generator of canonical evolution \cite{Jacobi1866}. In its modern form, the principle is standard in analytical mechanics \cite{Landau1976,Goldstein2002,Arnold1989,Lanczos1986}. Action stationarity is therefore not merely a method for deriving equations of motion; it is one of the first-principles languages of physical law. Beyond mechanics, it organizes variational formulations of electromagnetism, relativity, quantum theory, and field dynamics \cite{Lanczos1986,Yourgrau1979}. Through Noether's theorem, it also links continuous symmetries to conservation laws \cite{Noether1918}.

In its standard mechanical form, however, Hamilton's principle is not a bare statement of stationarity. It is stationarity under a boundary closure condition: admissible variations vanish at the endpoints of the variational interval. For the action functional
\[
S[q]
=
\int_{t_i}^{t_f}
L(q,\dot q,t)\,dt ,
\]
generated by the Lagrangian \(L(q,\dot q,t)\), the first variation separates into an interior term and a boundary term,
\[
\delta S
=
-
\int_{t_i}^{t_f}
\left[
\frac{d}{dt}
\left(
\frac{\partial L}{\partial \dot q}
\right)
-
\frac{\partial L}{\partial q}
\right]
\delta q\,dt
+
\mathcal B_{\partial}.
\]
Thus the interior term is governed by the Euler--Lagrange operator
\[
E[q]
\equiv
\frac{d}{dt}
\left(
\frac{\partial L}{\partial \dot q}
\right)
-
\frac{\partial L}{\partial q},
\]
whereas the boundary contribution is
\[
\mathcal B_{\partial}
=
[p\,\delta q]_{t_i}^{t_f},
\qquad
p
=
\frac{\partial L}{\partial \dot q},
\]
with \(p\) the canonical momentum \cite{Landau1976}. This term is not a disposable remainder: it is the variational interface between canonical momentum and endpoint admissibility.

The standard derivation imposes \(\delta q(t_i)=\delta q(t_f)=0\), hence \(\mathcal B_{\partial}=0\). Stationarity is then reduced to \(E[q]=0\), the on-shell Euler--Lagrange equation. Across the standard literature on analytical mechanics, fixed boundary admissibility is the natural choice for isolated deterministic systems \cite{Landau1976,Goldstein2002,Arnold1989,Lanczos1986}. The endpoints are prescribed, the admissible trajectory space is closed, and the variational boundary term vanishes. The closure condition therefore enters the formalism so early and so successfully that it is usually perceived as part of the variational method itself, rather than as a physical hypothesis about admissibility.

Contemporary variational mechanics often begins from a different situation. Statistical mechanics replaces the single closed trajectory by ensembles of many degrees of freedom, with macroscopic behaviour organized by energy, multiplicity, and coarse-grained microscopic motion \cite{Gibbs1902,Pathria2011}. Non-equilibrium statistical mechanics describes the effective influence of eliminated degrees of freedom \cite{Zwanzig2001,Bertin2013}; stochastic mechanics represents fluctuations through probabilistic evolution equations \cite{Risken1989}; and stochastic thermodynamics identifies irreversibility and entropy production as structural features of mesoscopic dynamics \cite{Seifert2012}, now accessible through fluctuation-based variance relations \cite{DiTerlizzi2024}. Even in quantum mechanics, where classical trajectories are no longer fundamental, the action remains the organizing object in variational \cite{Dirac1933} and path-integral formulations \cite{Feynman1948}. These frameworks extend mechanics beyond isolated motion, yet they usually operate after the variational problem has already been closed. This suggests returning one step earlier, to the boundary term itself.\\

The question is simple: \emph{what becomes of Hamilton's variational principle when the boundary part of the first variation is not required to vanish?}\\

The aim is not to modify the Lagrangian formalism or to introduce external forces by hand, but to make the closure assumption explicit and then relax it. We call this possibility \emph{variational openness}. Retaining the variational boundary term leads to an open Euler--Lagrange balance whose closed limit recovers the standard equation of motion. The paper develops this idea by isolating the closure assumption, defining variational openness, deriving the corresponding open Euler--Lagrange equation, and illustrating the construction through three elementary examples. It concludes by outlining an open Hamilton--Jacobi perspective, where openness acts not only on trajectories but on the action-generating structure itself.

\section{Results}

\subsection{The closure assumption in Hamilton's principle}
\label{sec:closure}

The standard derivation of the Euler--Lagrange equation begins from the action
\[
S[q]
=
\int_{t_i}^{t_f}
L(q,\dot q,t)\,dt .
\]

Its first variation is
\[
\delta S
=
\int_{t_i}^{t_f}
\left(
\frac{\partial L}{\partial q}\,\delta q
+
\frac{\partial L}{\partial \dot q}\,\delta \dot q
\right)dt .
\]
Integrating the second term by parts gives
\[
\delta S
=
\int_{t_i}^{t_f}
\left[
\frac{\partial L}{\partial q}
-
\frac{d}{dt}
\left(
\frac{\partial L}{\partial \dot q}
\right)
\right]
\delta q\,dt
+
\left[
\frac{\partial L}{\partial \dot q}\,\delta q
\right]_{t_i}^{t_f}.
\]

With \(p=\partial L/\partial\dot q\), the same result may be written as
\begin{equation}
\delta S
=
-
\int_{t_i}^{t_f}
\left[
\frac{d}{dt}
\left(
\frac{\partial L}{\partial \dot q}
\right)
-
\frac{\partial L}{\partial q}
\right]
\delta q\,dt
+
\mathcal B_{\partial},
\label{eq:R1-firstvariation}
\end{equation}
where
\[
\mathcal B_{\partial}
=
[p\,\delta q]_{t_i}^{t_f}
\]
is the variational boundary term.

Hamilton's principle requires stationarity, \(\delta S=0\). In the standard formulation, one imposes
\begin{equation}
\delta q(t_i)
=
\delta q(t_f)
=
0.
\label{eq:R1-fixed}
\end{equation}
Thus \(\mathcal B_{\partial}=0\), and the variational balance reduces to
\[
\int_{t_i}^{t_f}
\left[
\frac{d}{dt}
\left(
\frac{\partial L}{\partial \dot q}
\right)
-
\frac{\partial L}{\partial q}
\right]
\delta q\,dt
=
0.
\]

Since the interior variations remain arbitrary, the fundamental lemma of the calculus of variations gives
\begin{equation}
\frac{d}{dt}
\left(
\frac{\partial L}{\partial \dot q}
\right)
-
\frac{\partial L}{\partial q}
=
0.
\label{eq:R1-EL}
\end{equation}

The crucial point is that Eq.~(\ref{eq:R1-fixed}) is not a consequence of Hamilton's principle itself. It is an admissibility condition imposed before stationarity is evaluated. The familiar Euler--Lagrange equation therefore describes the variationally closed realization of Hamilton's principle, obtained after the boundary contribution has been removed.

\subsection{Variational openness}
\label{sec:openness}

The previous section showed that classical mechanics is obtained after imposing exact variational closure, \(\mathcal B_{\partial}=0\). We now consider the complementary possibility in which the boundary contribution is retained rather than removed. We define a variational problem as \emph{open} whenever
\begin{equation}
\mathcal B_{\partial}\neq0.
\label{eq:R2-open}
\end{equation}

The boundary term is the simplest representation of a more general variational fact: openness is carried by the boundary of the domain. In the mechanical realization, where the boundary consists of the two endpoints \(t_i\) and \(t_f\),
\[
\mathcal B_{\partial}
=
[p\,\delta q]_{t_i}^{t_f}
=
p(t_f)\,\delta q(t_f)
-
p(t_i)\,\delta q(t_i).
\]

Using the fundamental theorem of calculus, this quantity may be written as
\[
\mathcal B_{\partial}
=
\int_{t_i}^{t_f}
\frac{d}{dt}
\left(
p\,\delta q
\right)
dt.
\]

This motivates the introduction of a local boundary-openness density,
\begin{equation}
\rho_{\partial}(t)
\equiv
\frac{d}{dt}
\left(
p\,\delta q
\right),
\label{eq:R2-rho}
\end{equation}
so that
\[
\mathcal B_{\partial}
=
\int_{t_i}^{t_f}
\rho_{\partial}(t)\,dt.
\]

The quantity $\rho_{\partial}$ measures the local distribution of retained boundary influence throughout the variational domain. It is therefore a structural quantity associated with admissibility rather than a force, constitutive law, or dynamical model. Different physical realizations of variational openness may generate different forms of $\rho_{\partial}$, but the definition itself is entirely variational.

At this stage, $\rho_{\partial}$ still belongs to the variational balance because it contains the virtual displacement $\delta q$. The next step is therefore to determine how the retained boundary contribution enters the dynamical equation of motion.

\subsection{The open Euler--Lagrange equation}
\label{sec:oel}

The density $\rho_{\partial}$ still belongs to the variational balance because it contains the admissible variation $\delta q$. A dynamical equation requires a quantity that can multiply $\delta q$ as a source in the virtual-work balance. We therefore introduce a boundary-induced source $R_{\partial}$ through the weak identity
\begin{equation}
\int_{t_i}^{t_f}
\rho_{\partial}(t)\,dt
=
\int_{t_i}^{t_f}
R_{\partial}(t)\,
\delta q(t)\,dt .
\label{eq:R3-projection}
\end{equation}
Equivalently, $R_{\partial}$ is the representative whose virtual work reproduces the retained boundary contribution.

Substituting Eq.~(\ref{eq:R3-projection}) into the first variation transforms the retained boundary contribution into a source term acting on the admissible variations. Stationarity,
\[
\delta S=0,
\]
then gives the local balance
\begin{equation}
\frac{d}{dt}
\left(
\frac{\partial L}{\partial \dot q}
\right)
-
\frac{\partial L}{\partial q}
=
R_{\partial}(t).
\label{eq:R3-openEL}
\end{equation}

Equation~(\ref{eq:R3-openEL}) is the open Euler--Lagrange equation. The standard equation of motion is recovered in the exact-closure limit,
\[
R_{\partial}(t)=0.
\]

The construction therefore separates three objects that must not be identified,
\begin{equation}
\mathcal B_{\partial}
\longrightarrow
\rho_{\partial}
\longrightarrow
R_{\partial},
\label{eq:R3-chain}
\end{equation}
namely the retained boundary term, its variational density, and its projected dynamical representative. The projection is generally not unique: different physical realizations of the same retained boundary contribution may correspond to different representatives within the same variational class. The examples below illustrate several elementary realizations of this construction.

\subsection{From boundary openness to dynamical sources}
\label{sec:boundarytosource}

The projected source \(R_{\partial}\) is not introduced as an arbitrary force. It is a dynamical representative of a retained boundary contribution. In the mechanical realization,
\[
\mathcal B_{\partial}
=
p(t_f)\,\delta q(t_f)
-
p(t_i)\,\delta q(t_i).
\]
Thus openness measures a mismatch between canonical momentum and admissible endpoint variation.

A projected realization is obtained when the retained boundary contribution is represented as time-distributed virtual work,
\begin{equation}
\mathcal B_{\partial}
=
\int_{t_i}^{t_f}
R_{\partial}(t)\,\delta q(t)\,dt .
\label{eq:R4-virtualwork}
\end{equation}
Equation~(\ref{eq:R4-virtualwork}) is not an additional force law. It is a weak representation of the retained boundary term on the admissible variations. Different boundary preparations can therefore produce different representatives \(R_{\partial}\) within the same variational class. The examples below specify elementary choices of this representative.

\subsection{The open harmonic oscillator}
\label{sec:oscillator}

Consider the harmonic oscillator with Lagrangian
\begin{equation}
L(q,\dot q)
=
\frac{1}{2}
m\dot q^2
-
\frac{1}{2}
kq^2 .
\label{eq:R5-L}
\end{equation}

Since \(\partial L/\partial \dot q=m\dot q\) and \(\partial L/\partial q=-kq\), the open Euler--Lagrange equation becomes
\begin{equation}
m\ddot q
+
kq
=
R_{\partial}(t).
\label{eq:R5-openosc}
\end{equation}

To display the effect explicitly, consider the minimal case in which the projected boundary source is constant, \(R_{\partial}(t)=R_0\). Then
\[
m\ddot q
+
kq
=
R_0,
\]
with solution
\begin{equation}
q(t)
=
A\cos(\omega_0 t)
+
B\sin(\omega_0 t)
+
\frac{R_0}{k},
\qquad
\omega_0^2
=
\frac{k}{m}.
\label{eq:R5-shiftedsolution}
\end{equation}

Thus a constant projected boundary mismatch shifts the equilibrium position by \(R_0/k\) without changing the intrinsic oscillation frequency. In the exact-closure limit \(R_{\partial}(t)=0\), Eq.~(\ref{eq:R5-openosc}) reduces to the standard harmonic oscillator,
\[
m\ddot q
+
kq
=
0.
\]

\subsection{Finite boundary compliance}
\label{sec:compliance}

A physically transparent realization of variational openness is obtained by replacing a perfectly fixed endpoint with a boundary of finite stiffness. Consider a final boundary action
\begin{equation}
S_{\partial}^{(f)}
=
\frac{\kappa}{2}
\left(
q_f-Q_f
\right)^2 ,
\label{eq:R6-Sboundary}
\end{equation}
where \(q_f=q(t_f)\), \(Q_f\) is the prescribed endpoint, and \(\kappa\) is the boundary stiffness.

Its variation is
\[
\delta S_{\partial}^{(f)}
=
\kappa
\left(
q_f-Q_f
\right)
\delta q_f .
\]
The total final boundary contribution is therefore
\[
\mathcal B_{\partial}^{(f)}
=
p_f\,\delta q_f
+
\kappa
\left(
q_f-Q_f
\right)
\delta q_f
=
\left[
p_f
+
\kappa
\left(
q_f-Q_f
\right)
\right]
\delta q_f ,
\]
where \(p_f=p(t_f)\).

The natural closed-boundary condition associated with the total variational problem is
\begin{equation}
p_f
+
\kappa
\left(
q_f-Q_f
\right)
=
0.
\label{eq:R6-naturalBC}
\end{equation}

If this condition is not imposed exactly, the residual boundary mismatch defines a localized projected source. In distributional form,
\begin{equation}
R_{\partial}(t)
=
\kappa
\left(
q_f-Q_f
\right)
\delta(t-t_f).
\label{eq:R6-source}
\end{equation}

This example shows that variational openness need not be stochastic and need not be imposed by hand as a bulk force. It can arise from a finite boundary action. In the limit \(\kappa\rightarrow\infty\), the endpoint is effectively fixed and standard endpoint closure is recovered. For finite \(\kappa\), the endpoint remains partially admissible, and the variational problem is only partially closed.

\subsection{Memory and delayed openness}
\label{sec:delay}

The retained boundary contribution need not be localized at a single endpoint. It may be distributed over the history of the admissible path. A minimal way to represent this situation is through a non-local openness action,
\begin{equation}
S_{\partial}^{\rm mem}
=
\frac{1}{2}
\int_{0}^{t}
\int_{0}^{t}
q(u)\,
K(u-s)\,
q(s)\,du\,ds ,
\label{eq:R7-Smemory}
\end{equation}
where \(K(u-s)\) is a memory kernel.

Assuming a symmetric kernel \(K(u-s)=K(s-u)\), its variation is
\[
\delta S_{\partial}^{\rm mem}
=
\int_{0}^{t}
\left[
\int_{0}^{t}
K(u-s)\,
q(s)\,ds
\right]
\delta q(u)\,du .
\]
The corresponding projected source is therefore
\begin{equation}
R_{\partial}(u)
=
\int_{0}^{t}
K(u-s)\,
q(s)\,ds .
\label{eq:R7-source-memory}
\end{equation}

For the open harmonic oscillator, this gives
\[
m\ddot q(t)
+
kq(t)
=
\int_{0}^{t}
K(t-s)\,
q(s)\,ds .
\]

A simple delayed realization is obtained by choosing
\[
K(t-s)
=
\alpha\,
\delta(t-s-\tau),
\]
which gives
\begin{equation}
m\ddot q(t)
+
kq(t)
=
\alpha q(t-\tau).
\label{eq:R7-delay}
\end{equation}

Thus memory and delay are not introduced as independent dynamical postulates. They arise here as projected consequences of a retained non-local openness action.

\subsection{Open Hamilton--Jacobi perspective}
\label{sec:hj}

The preceding construction acts directly on the variational structure and therefore suggests a corresponding extension of Hamilton--Jacobi theory. To avoid confusion with the action functional \(S[q]\), we denote Hamilton's principal function by \(W(q,t)\).

For a variationally closed system, Hamilton's principal function satisfies
\begin{equation}
H
\left(
q,
\frac{\partial W_0}{\partial q},
t
\right)
+
\frac{\partial W_0}{\partial t}
=
0.
\label{eq:R8-HJ}
\end{equation}

The open variational balance developed above can be written schematically as
\[
\delta S_{\mathrm{bulk}}
+
\delta S_{\partial}
=
0,
\]
suggesting that openness may also be represented at the level of the action-generating function itself.

The corresponding generalized Hamilton--Jacobi equation takes the form
\begin{equation}
H
\left(
q,
\frac{\partial W}{\partial q},
t
\right)
+
\frac{\partial W}{\partial t}
=
\Sigma_{\partial},
\label{eq:R8-openHJ}
\end{equation}
where \(\Sigma_{\partial}\) denotes an action-level source associated with retained variational openness.

Equation~(\ref{eq:R8-openHJ}) should be regarded as a structural extension rather than a completed theory. Its significance is that openness acts directly on the action-generating function, and therefore on the variational structure from which the equations of motion are derived.

\subsection{Weak variational openness}
\label{sec:perturbation}

When the retained boundary contribution is small, openness can be treated perturbatively. Let \(\varepsilon\ll1\) measure the departure from exact variational closure. The Hamilton principal function is then expanded as
\begin{equation}
W
=
W_0
+
\varepsilon W_1
+
O(\varepsilon^2),
\label{eq:R9-actionexpansion}
\end{equation}
where \(W_0\) generates the closed dynamics and \(W_1\) is the leading correction induced by openness.

The action-level source is expanded consistently as
\begin{equation}
\Sigma_{\partial}
=
\varepsilon \Sigma_{\partial}^{(1)}
+
O(\varepsilon^2).
\label{eq:R9-sourceexpansion}
\end{equation}

Substitution into Eq.~(\ref{eq:R8-openHJ}) gives, at zeroth order, the closed Hamilton--Jacobi equation,
\[
H
\left(
q,
\frac{\partial W_0}{\partial q},
t
\right)
+
\frac{\partial W_0}{\partial t}
=
0.
\]

At first order, the correction satisfies the forced transport equation
\begin{equation}
v_H(q,t)\,
\frac{\partial W_1}{\partial q}
+
\frac{\partial W_1}{\partial t}
=
\Sigma_{\partial}^{(1)},
\label{eq:R9-order1}
\end{equation}
where
\[
v_H(q,t)
\equiv
\left.
\frac{\partial H}{\partial p}
\right|_{p=\partial_q W_0}
\]
is the Hamiltonian velocity field evaluated on the closed solution.

Equivalently,
\[
\mathcal L_H[W_1]
=
\Sigma_{\partial}^{(1)},
\]
where \(\mathcal L_H\) is the Hamilton--Jacobi transport operator linearized around \(W_0\).

Thus weak variational openness appears as a perturbation of the action-generating structure itself. In the limit \(\varepsilon\rightarrow0\), the standard closed Hamilton--Jacobi theory is recovered.

\section{Discussion}

The central result of this work is that the standard Euler--Lagrange equation is not the most general consequence of Hamilton's principle. It is obtained only after imposing exact variational closure through the elimination of the boundary contribution. Once this assumption is relaxed, the variational principle admits an open realization in which the familiar Euler--Lagrange dynamics appears as the exactly closed limit.

The proposed construction does not modify the Lagrangian formalism and does not introduce external forces by postulate. Instead, it identifies source terms with the dynamical imprint of incomplete variational closure. In this interpretation, forcing, finite boundary compliance, memory kernels, and delayed responses are different realizations of retained boundary influence. The source is not added to the equation of motion from outside; it emerges from relaxing the admissibility conditions under which that equation is derived.

The examples developed above illustrate three elementary forms of the same mechanism. The open harmonic oscillator shows how a projected boundary mismatch appears as an effective source. The finite-compliance boundary shows that openness can arise from a primitive boundary action and imperfect preparation rather than from stochastic forcing. The delayed oscillator shows that retained openness can be distributed over the history of the system, producing memory and non-Markovian response. Together, these examples realize the central variational chain
\[
\mathcal B_{\partial}
\longrightarrow
\rho_{\partial}
\longrightarrow
R_{\partial}
\longrightarrow
E[q]=R_{\partial},
\]
through which retained boundary admissibility becomes effective dynamics.

The framework is structural rather than constitutive. The density \(\rho_{\partial}\) is a variational object, not a force; the projected source \(R_{\partial}\) is not a universal constitutive law; and, in general, the projection from \(\mathcal B_{\partial}\) to \(R_{\partial}\) is not unique. Different physical realizations may correspond to different dynamical representatives of the same retained variational contribution. Establishing existence and uniqueness criteria for this projection is a mathematical problem beyond the scope of the present work. The present result therefore identifies where openness enters the variational principle, not which microscopic mechanism must generate it.

The Hamilton--Jacobi extension points toward a deeper level of the same structure. If openness acts on the action-generating function itself, then deviations from exact closure may be organized as perturbations of the action rather than only as perturbations of trajectories. In this sense, the closed Hamilton--Jacobi equation appears as the zeroth-order limit of a weakly open action theory.

A further consequence concerns Noether's theorem. In variationally closed systems, continuous symmetries generate exactly conserved currents. Since variational openness replaces the closed balance by
\[
E[q]
=
R_{\partial},
\]
the associated conservation laws are expected to acquire source terms proportional to the failure of exact closure. Schematically,
\[
\frac{dJ}{dt}
=
Q_{\partial},
\]
or, in field-theoretic language,
\[
\partial_\mu J^\mu
=
Q_{\partial},
\]
where \(Q_{\partial}\) measures the openness-induced departure from exact conservation. From this perspective, variational openness may be interpreted as a controlled departure from exact Noether conservation. Developing a complete open-Noether theorem remains an interesting problem for future work.

The construction also has a natural field-theoretic analogue. In Maxwell electrodynamics, the action
\[
S_{\rm EM}
=
-\frac{1}{4}
\int
F_{\mu\nu}F^{\mu\nu}\,d^4x
\]
has a variation containing both bulk and boundary contributions. In the closed variational problem, the bulk equation is
\[
\partial_{\mu}F^{\mu\nu}=0.
\]
If the boundary contribution is retained and projected onto admissible variations of the gauge potential, the same variational logic gives an open balance of the form
\[
\partial_{\mu}F^{\mu\nu}
=
J_{\partial}^{\nu},
\]
where \(J_{\partial}^{\nu}\) is the effective current associated with incomplete variational closure. This does not claim that electromagnetic charge is reducible to a boundary term. It states that the same structural mechanism by which a retained boundary contribution becomes a source in mechanics also appears in a canonical field theory. In this Abelian case, variational openness can be read as an effective flux imbalance carried by the boundary of the electromagnetic domain.

\section{Outlook}

The first extension beyond Maxwell electrodynamics is to non-Abelian gauge theory. For Yang--Mills fields, the action
\[
S_{\rm YM}
=
-\frac{1}{4}
\int
F^a_{\mu\nu}F_a^{\mu\nu}\,d^4x
\]
leads, under closed variational admissibility, to the bulk equation
\[
(D_{\mu}F^{\mu\nu})_a=0,
\]
where \(D_\mu\) is the gauge-covariant derivative. An open variational balance would formally produce
\[
(D_{\mu}F^{\mu\nu})_a
=
(J_{\partial})^\nu_a,
\]
where \((J_{\partial})^\nu_a\) denotes a gauge-valued effective current associated with retained boundary admissibility. Unlike the Abelian Maxwell case, variational openness is now constrained by the nonlinear gauge structure. The effective source carries an internal gauge index and therefore represents not only incomplete closure across spacetime boundaries, but also incomplete closure within the gauge-orbit structure of the theory. Since non-Abelian gauge currents are themselves associated with symmetry generators, variational openness would be expected to induce controlled departures from exact gauge-current conservation, providing a natural extension of the open-Noether perspective outlined above. This formulation is only schematic, but it identifies non-Abelian gauge theory as a natural arena for testing whether incomplete variational closure can be made compatible with covariance, gauge symmetry, and source structure.

A quantum example is equally instructive. In a variational formulation of Schrödinger dynamics, the wavefunction replaces the classical trajectory, but the action remains the organizing object. A closed variational problem yields
\[
i\hbar
\frac{\partial \psi}{\partial t}
=
\hat H\psi .
\]
An open variational balance would have the schematic form
\[
i\hbar
\frac{\partial \psi}{\partial t}
=
\hat H\psi
+
\Sigma_{\partial}[\psi],
\]
where \(\Sigma_{\partial}[\psi]\) denotes the variational imprint of retained admissibility in the quantum action. This equation is not intended as a complete theory of open quantum systems; it only indicates how the present construction may survive when classical trajectories are no longer fundamental. In this setting, quantities that are exactly conserved in the closed theory may acquire openness-induced source terms. Variational openness therefore suggests a route by which departures from exact conservation could emerge directly at the level of the quantum action rather than through phenomenological modifications of the dynamics.

Gravitation provides a sharper and more delicate test. The Einstein--Hilbert action is famously sensitive to boundary terms, and the Gibbons--Hawking--York contribution is required for a well-posed variational problem. From the present perspective, such boundary terms may be more than technical completions: they may encode the conditions under which geometry itself is variationally closed or open. A generally covariant theory of variational openness remains to be formulated.

A further direction concerns statistical and quantum open systems. Coarse-graining, eliminated degrees of freedom, memory kernels, and dissipative terms are usually introduced after the dynamical equations have been written. Variational openness suggests a complementary route: some effective open dynamics may arise from a variational problem whose boundary admissibility was not exactly closed.

Finally, the open Hamilton--Jacobi perspective suggests possible applications to emergent spacetime and biological action organization. In cosmological settings, openness may become relevant in regimes where geometry, boundary admissibility, and causal structure are not yet cleanly separated. In such regimes, openness would not merely modify dynamics within spacetime; it could participate in defining the admissibility of the variational domain itself. In this view, causal structure and admissible geometry could emerge simultaneously from a variational selection process, with exact variational closure appearing only after the admissible spacetime domain has become dynamically established. In biological systems, weakly open action-generating structures,
\[
W
=
W_0
+
\varepsilon W_1
+
O(\varepsilon^2),
\]
may provide a framework in which stable functional modes emerge as dynamically selected sectors of incomplete variational closure. These possibilities are not developed here, but they motivate future work on systems in which action, admissibility, coherence, and stability are inseparable.

\section{Conclusion}

Hamilton's principle is usually applied after fixing the admissible variations at the boundary. This work has identified that step as a closure hypothesis rather than as a merely technical convention. Removing the boundary contribution yields the standard Euler--Lagrange equation; retaining it leads, within the same variational principle, to an open Euler--Lagrange balance.

The retained boundary term defines a variational density of openness. After projection onto admissible variations, this density becomes a boundary-induced dynamical source. The exact-closure limit \(R_{\partial}=0\) recovers the usual equation of motion. The examples show that forcing, finite boundary compliance, memory, and delay can be understood as distinct realizations of incomplete variational closure.

The framework therefore does not replace Hamiltonian mechanics; it enlarges the admissibility structure under which Hamilton's principle is applied. Open dynamics need not be formulated only by adding forces to closed equations. It may also be formulated by relaxing the closure assumptions from which those equations are obtained. From this perspective, Hamiltonian mechanics appears not as the variational theory of all dynamical systems, but as the special case in which variational closure is exact.

\backmatter

\bmhead{Acknowledgements}

The author acknowledges discussions with colleagues on analytical mechanics, open systems, and variational principles.

\section*{Declarations}

\bmhead{Funding}
Not applicable.

\bmhead{Conflict of interest}
The author declares no competing interests.

\bmhead{Ethics approval and consent to participate}
Not applicable.

\bmhead{Consent for publication}
Not applicable.

\bmhead{Data availability}
No datasets were generated or analyzed during the current theoretical study.

\bmhead{Code availability}
Not applicable.

\bmhead{Author contribution}
Francisco Monroy conceived and wrote the manuscript.

\bibliography{refs}

\end{document}